\documentclass[prb,twocolumn,superscriptaddress, nobibnotes]{revtex4}
\usepackage{amsfonts}
\usepackage{epsfig}
\usepackage{bm}
\usepackage{amsmath}
\usepackage{graphicx}
\usepackage{float}
\usepackage{hyperref}
\hypersetup{colorlinks=true}

\begin{document}

\title{Field-induced insulating states in a graphene superlattice}

\author{S. Pezzini}
\altaffiliation{Present address: Center for Nanotechnology Innovation @ NEST, Italian Institute of Technology, P.za S. Silvestro 12, 56127 Pisa, Italy; sergio.pezzini@iit.it}
\affiliation{High Field Magnet Laboratory (HFML-EMFL), Radboud University, NL-6525 ED Nijmegen, The Netherlands}
\affiliation{Institute for Molecules and Materials, Radboud University, NL-6525 AJ Nijmegen, The Netherlands}

\author{S. Wiedmann}
\affiliation{High Field Magnet Laboratory (HFML-EMFL), Radboud University, NL-6525 ED Nijmegen, The Netherlands}
\affiliation{Institute for Molecules and Materials, Radboud University, NL-6525 AJ Nijmegen, The Netherlands}

\author{A. Mishchenko}
\affiliation{School of Physics and Astronomy, University of Manchester, Oxford Road, Manchester M13 9PL, UK}

\author{ M. Holwill}
\affiliation{School of Physics and Astronomy, University of Manchester, Oxford Road, Manchester M13 9PL, UK}

\author{R. Gorbachev}
\affiliation{School of Physics and Astronomy, University of Manchester, Oxford Road, Manchester M13 9PL, UK}

\author{D. Ghazaryan}
\affiliation{School of Physics and Astronomy, University of Manchester, Oxford Road, Manchester M13 9PL, UK}
\affiliation{Department of Physics, National Research University Higher School of Economics,Staraya Basmannaya 21/4, Moscow 105066, Russian Federation}

\author{K. S. Novoselov}
\affiliation{School of Physics and Astronomy, University of Manchester, Oxford Road, Manchester M13 9PL, UK}

\author{U. Zeitler}
\affiliation{High Field Magnet Laboratory (HFML-EMFL), Radboud University, NL-6525 ED Nijmegen, The Netherlands}
\affiliation{Institute for Molecules and Materials, Radboud University, NL-6525 AJ Nijmegen, The Netherlands}

\begin{abstract}
We report on high-field magnetotransport ($B$ up to 35 T) on a gated superlattice based on single-layer graphene aligned on top of hexagonal boron nitride. The large-period moir\'e modulation ($\approx 15$ nm) enables us to access the Hofstadter spectrum in the vicinity of and above one flux quantum per superlattice unit cell ($\mathit{\Phi/\Phi_0}=1$ at $B = 22$ T). We thereby reveal, in addition to the spin-valley antiferromagnet at $\nu=0$, two insulating states developing in positive and negative \textit{effective} magnetic fields from the main $\nu=1$ and $\nu=-2$ quantum Hall states respectively. We investigate the field dependence of the energy gaps associated with these insulating states, which we quantify from the temperature-activated peak resistance. Referring to a simple model of local Landau quantization of third generation Dirac fermions arising at $\mathit{\Phi/\Phi_0}=1$, we describe the different microscopic origins of the insulating states and experimentally determine the energy-momentum dispersion of the emergent gapped Dirac quasi-particles.
\end{abstract}
\maketitle

\section{Introduction}
Van der Waals assembly of atomically-thin materials represents a novel powerful strategy for the realization of artificial structures with tailored electronic response.\cite{Geim_13} Inherent to this approach is the control of the crystallographic orientation of the atomic layers, a novel degree of freedom that can profoundly alter the electrostatic landscape experienced by the charge carriers. The stack of graphene \cite{Novoselov_04} on top of hexagonal boron nitride (hBN),\cite{Dean_10} and the subsequent reconstruction of the electronic spectrum hosting 'second generation' Dirac cones,\cite{Yankowitz_12} is a prototypical case. The small lattice mismatch ($\approx1.8$\%) between their isomorphic structures results into an hexagonal superlattice modulation (so-called moir\'e pattern),\cite{Xue_11, Decker_11,Yankowitz_12} which sets an artificial periodicity as large as $\lambda\approx15$ nm for perfect crystallographic alignment. This twist-dependent superstructure, combined with the possibility of in-situ tuning of the band filling via electrostatic field-effect, has made graphene-hBN superlattices the ideal platform for the experimental study of the Hofstadter butterfly (HB).\cite{Hofstadter_76,Ponomarenko_13,Dean_13,Hunt_13} The HB is the fractal (i.e. recursive, self-similar) energy spectrum acquired by a two-dimensional (2D) electronic system when simultaneously subjected to (i) a periodic electrostatic potential (the hexagonal moir\'e in our case) and (ii) a perpendicular magnetic field.\cite{Hofstadter_76} A spatially periodic potential groups the electronic states into discrete Bloch bands;\cite{Ashcroft} a perpendicular magnetic field quantizes the electronic spectrum into Landau levels.\cite{Kittel} These fundamental effects can usually be treated independently, however under particular circumstances the two quantizations combine into the HB. This happens when rational values of magnetic flux quanta ($\Phi_0=h/e$) threads the superlattice unit cell, i.e. when $\mathit{\Phi/\Phi_0}=BA/(h/e)=p/q$ (where $\Phi$ is the total magnetic flux, $A$ is the superlattice unit cell area, $h$ the Planck constant, $e$ the electron's charge). Thereby, the Bloch (Landau) bands splits into $q$ ($p$) sub-bands, leading to a repeated 'cloning' of the original magnetic spectrum.\cite{Hofstadter_76}\\
In the case of graphene superlattices, the HB combines with the specific response of graphene's 2D Dirac fermions to large magnetic fields.\cite{Novoselov_05, Zhang_05} When tuned to charge neutrality, graphene systems exhibits a field-induced insulating state which has been a subject of intensive experimental study in single-layer,\cite{Zhang_06, Checkelsky_07, Giesbers_09, Du_09, Yu_13, Young_14} bilayer\cite{Feldman_09, Freitag_12, Velasco_12, Maher_13, Pezzini_14} and multi-layers.\cite{Bao_11, Grushina_15} This state arises at half filling of the zero-energy Landau level (LL) - a unique signature of Dirac fermions - and has an interaction-induced origin based on the so-called quantum Hall (QH) ferromagnetism (QHFM), i.e. on the formation of spin-valley polarized states at partial LL fillings.\cite{Young_12} It is therefore of fundamental interest to understand if 'copies' of this state are present in graphene's HB, and in which way their phenomenology does differ from the 'original' one. Using capacitance spectroscopy Yu \textit{et al.} already showed evidence for QHFM in graphene superlattices.\cite{Yu_14} However, magneto-capacitance is sensitive only to the bulk density of states and does not allow discerning between QH and insulating states. On the other hand, electrical transport can be employed to identify insulating phases with an energy gap for both the bulk and edge excitations. Bearing this in mind, we have investigated a 15 nm graphene-hBN superlattice with high-field temperature-dependent magnetotransport measurements. By this means, we reveal three field-induced insulating states in the HB and quantitatively estimate their activation energies and the gaps' field dependence. Their microscopic origin is interpreted in connection to the emergence of so-called 'third generation' Dirac particles, i.e. field-and-superlattice-induced replica, experiencing zero effective magnetic field when $\mathit{\Phi=\Phi_0}$ (i.e. at $B=22$ T for our superlattice). By analyzing their local Landau quantization, we reveal a significant renormalization in the Fermi velocity of the corresponding gapped Dirac cones.

\begin{figure}[!t]
\centerline{\includegraphics[width = 0.373\textwidth]{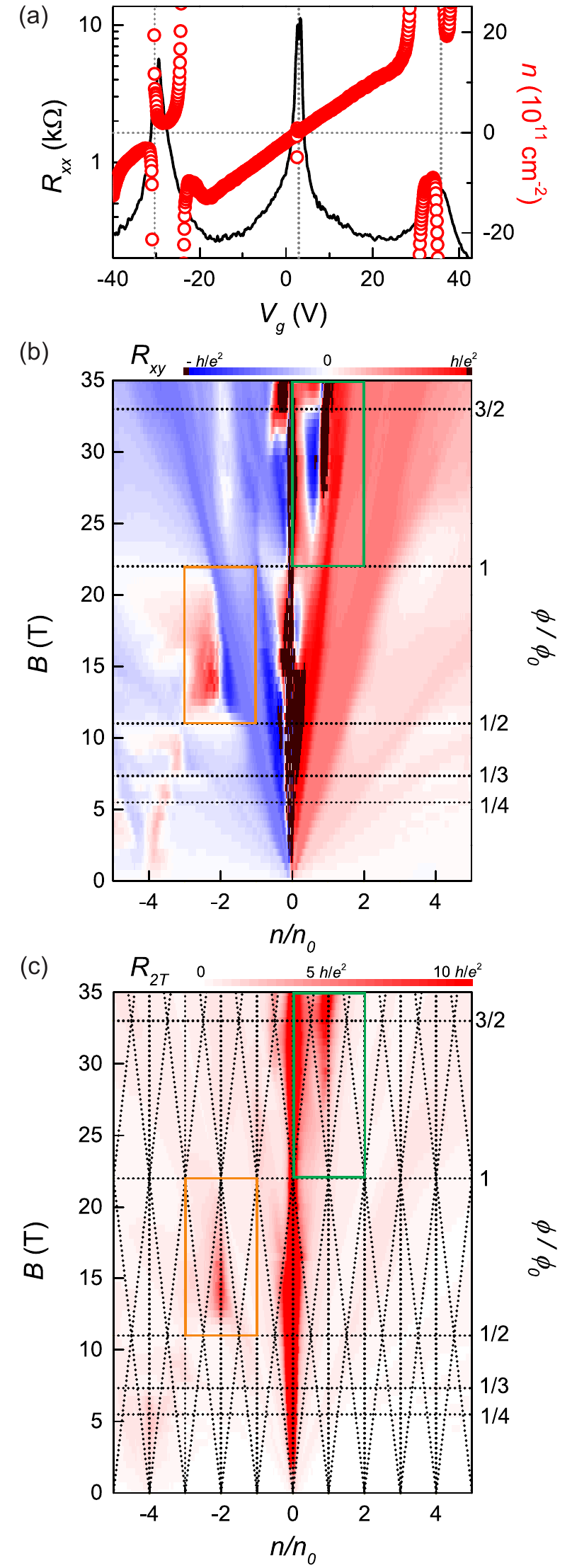}}
\caption{(a) Black line: $R_{xx}$($V_g$) at $B=0$ T. Red circles: $n$($V_g$), extracted from $R_{xy}$ at $B=0.5$ T. The vertical dotted lines indicate the main and satellite CNPs. (b) $R_{xy}$ as a function of $B$ and charge density per superlattice unit cell. On the right axis $B$ is converted to $\mathit{\Phi/\Phi_0}$, with some rational values indicated. (c) $R_{2T}$ over the same field and density ranges as in (b). The vertical (diagonal) dotted lines correspond to gaps in the Wannier diagram with index $t=0$ ($t=\pm1$). The maps were acquired as $V_g$ sweeps at constant $B$, with 0.5 T steps; $T=1.5$ K, unless otherwise specified. The green and orange rectangles mark the two regions discussed in Fig.\ref{Two}.}
\label{One}
\end{figure}

\section{Experimental methods}
We studied a six-terminal Hall bar device ($W\times L = 1 \times 2$ $\mathrm{\mu}$m$^2$), defined by reactive ion etching and evaporation of Cr/Au contacts. The sample is based on a stack of single-layer graphene on top of a 50 nm thick hBN crystal, obtained with a dry van der Waals assembly technique.\cite{Kretinin_14} Straight edges of the two crystals are aligned within $\approx1^{\circ}$ during the assembly (giving a 50\% success chance of crystallographic alignment due to the uncertainty on the zig-zag or armchair nature of the selected edges). The measurements presented were performed in a $^4$He-flow cryostat (base temperature 1.5 K), inserted in the access bore of a resistive Bitter magnet at the High Field Magnet Laboratory (HFML-EMFL). The resistance was measured with low-frequency lock-in detection, both in two and four-probe configuration, with a constant ac voltage of 10 mV applied to the sample connected in series to a 100 k$\Omega$ resistor.

\section{Results and Discussion}
\subsection{Hofstadter butterfly}
\begin{figure*}[!t]
\centerline{\includegraphics[width =\textwidth]{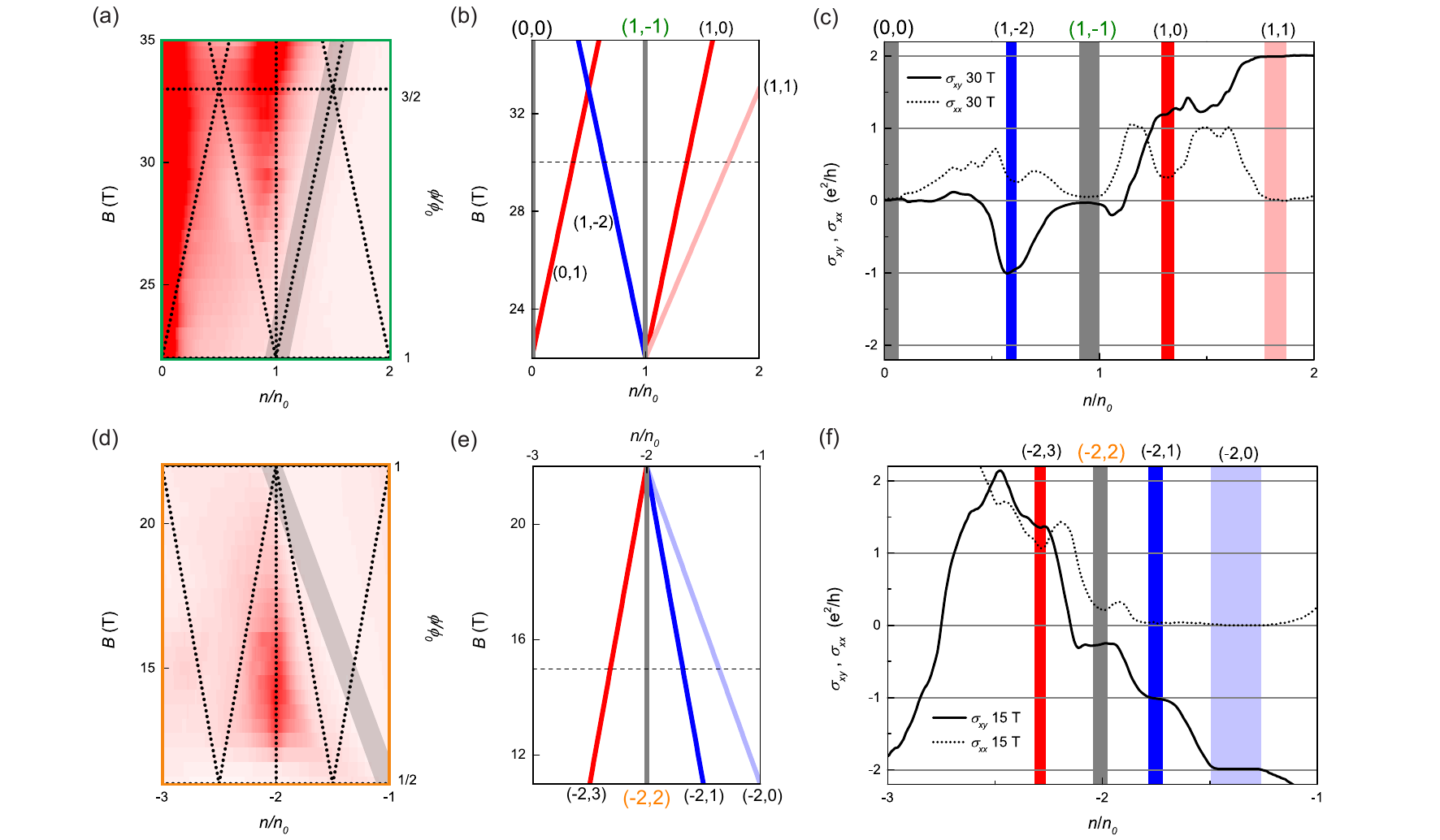}}
\caption{(a,d) Enlarged plots of $R_{2T}$ in the regions hosting the insulating states (1,-1) and (-2,2) (green and orange rectangles in Fig.\ref{One} (b),(c)). The shaded areas correspond to the parent QH states, $\nu=1$ and $\nu=-2$ respectively. (b,e) Local QH fan for the two regions in (a,d), labelled according to the notation ($\nu$,$\nu_L$).\cite{Yu_14} The colours indicate the expected Hall conductivity $\sigma_{xy}[(e^2/h)]=\nu+\nu_L$ (grey $=0$, blue $=-1$, red $=+1$, light blue $=-2$, light red $=+2$). (c,f) $\sigma_{xx}$ and $\sigma_{xy}$ at the field indicated by the dashed lines in (b),(e), over the same density range as in the other panels. The local QH states are highlighted by the same colour-code as in the diagrams of panels (b),(e).} 
\label{Two}
\end{figure*}
The black trace in Fig.\ref{One} (a) shows the four terminal resistance of our device ($R_{xx}$) as a function of gate voltage ($V_g$, applied to the underlying Si/SiO$_2$ substrate), measured at $B=0$ T and $T=1.5$ K. As expected for an aligned graphene-hBN stack, three pronounced peaks are visible. At $V_g=V_g^{CNP}\approx3$ V the Fermi level is set at the touching between the conduction and valence band Dirac cones. The peak value of the resistance at this point remains $\approx 10$ k$\Omega$ (slightly varying over different cooldowns), indicating that no appreciable band-gap opens in the absence of a magnetic field. The two additional maxima symmetrically located at large positive (negative) doping $|\Delta V_g |= |V_g - V_g^{CNP}|= 32.5$ V, on the other hand, correspond to half filling of \textit{second generation} Dirac mini-bands induced by the superlattice potential.\cite{Yankowitz_12} To confirm this identification, we measured the Hall resistance $R_{xy}$ as a function of $V_g$ at low magnetic field ($B=0.5$ T, avoiding quantization effects) and extracted the corresponding 2D carrier density $n(V_g)=B/(eR_{xy})$ (red open circles in Fig.\ref{One} (a)). Close to $V_g=0$ V, $n$ changes its sign as the system is set to the charge neutrality point (CNP), while it varies linearly with $V_g$ otherwise, with a slope $\alpha=6.3\times10^{10}$ cm$^{-2}$/V. Further away from the main CNP, $n$ changes its sign a first time at the onset of the superlattice mini-bands ($| \Delta V_g |= 26$ V, corresponding to van Hove singularities,\cite{Wallbank_13}) and a second time at $|\Delta V_g |=32.5$ V, indicating the CNPs for the superlattice-induced Dirac mini-bands. Due to the spin and valley degeneracy, these satellite CNPs are realized when accommodating four electrons per superlattice unit cell, i.e. when $n=4n_0=2.05\times10^{12}$ cm$^{-2}$, with $n_0=1/A=2/(\sqrt{3}\lambda^2)$. The positioning of the satellite peaks therefore allows to estimate a periodicity $\lambda=15$ nm for the hexagonal moir\'e pattern, which confirms the high degree of crystallographic alignment between the graphene and hBN crystals.
\\In Fig.\ref{One} (b) we present a colour map of the Hall resistance as function of magnetic field (0 T $<B<35$ T) and carrier density per superlattice unit cell $n/n_0$ (corresponding to the same range of $V_g$ used in (a)). The red areas indicate electron doping, the blue ones hole doping, while in the black ones $|R_{xy}|$ exceeds $h/e^2$, typically signalling a divergence due to a CNP. On top of the standard Landau fan diagram of single-layer graphene (with full degeneracy lifting of the $N=0$ LL), one can clearly identify a recursive pattern due to the HB. This is particularly evident in the lower-left part of the map, which is dominated by a series of charge conversions and local Landau mini-fans. These features are understood in terms of the formation of $q$-fold degenerate Zak mini-bands,\cite{Zak_64} which are in fact gapped Dirac cones,\cite{Chen_14} and experience zero effective field $B_{eff}$ when $\mathit{\Phi/\Phi_0}=p/q$. The emergent \textit{third generation} Dirac quasi-particles are then subjected to Landau quantization in a finite (positive or negative) effective magnetic field $B_{eff}=B-\mathit{\Phi_0}A p/q$. The appearance of these structures have a clear $1/B$ periodicity, with a characteristic frequency $f=\mathit{\Phi_0}/A=22$ T, that provides an alternative way to estimate the moir\'e periodicity $\lambda=14.7$ nm, in reasonable accordance with the value given above.
\begin{figure*}[!t]
\centerline{\includegraphics[width =\textwidth]{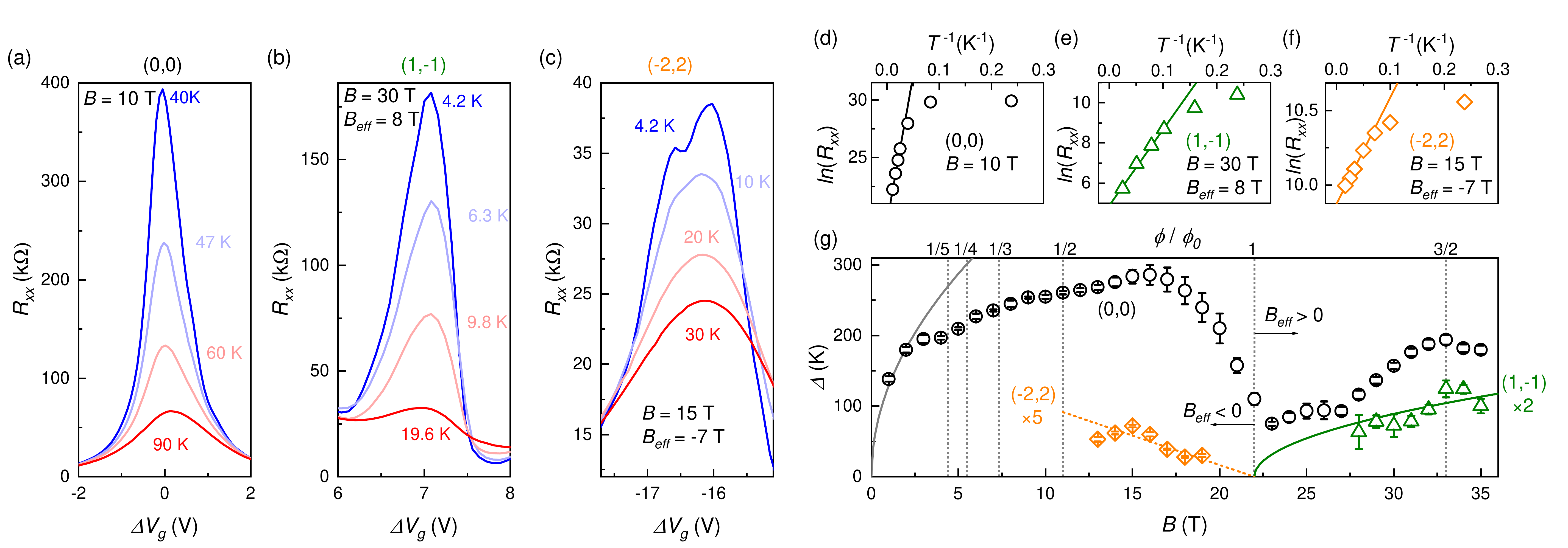}}
\caption{(a-c) $R_{xx}(V_g)$ at selected temperatures in correspondence of the three insulating states at representative values of the magnetic field. (d-f) Maxima of $R_{xx}$ (in ln scale) as a function of $1/T$, extracted from $V_g$ sweeps at constant $B$, such as the ones shown in (a-c). The solid lines are fit to the activation law $\ln(R_{xx})\alpha \Delta/2k_BT$. The resulting energy gaps $\Delta$ are plotted in (g) in Kelvin units, as a function of the magnetic field. The grey line represents the Coulomb energy, the green continuous one is a $\sqrt{B_{eff}}$-fit to $\Delta$(1,-1), while the orange dashed one is a linear fit to $\Delta$(-2,2) for $B_{eff}<0$. The data and fits for $\Delta$(-2,2) and $\Delta$(1,-1) are multiplied by a factor 5 and 2, respectively.}
\label{Three}
\end{figure*}

\subsection{Insulating regions within local Landau fans}
Within the HB, a simple Diophantine relation $n/n_0=t(\mathit{\Phi/\Phi_0})+s$ (with $t$, $s$ integer numbers, although fractional indices were recently reported in Refs.\onlinecite{Wang_15,Spanton_17}) locates the expected incompressible (i.e. bulk-gapped) states in the flux-density space. The slope $t$ can be seen as a generalized filling factor and actually defines the expected Hall conductivity $\sigma_{xy}=t(e^2/h)$, while $s$ indicates the amount of filling of the Bloch bands. In Fig.\ref{One} (c) we plot a grid of $n/n_0=t(\mathit{\Phi/\Phi_0})+s$ lines, i.e. a so-called Wannier diagram,\cite{Wannier_78} limited to $t=0,\pm1$ only, on top of a color map of the two-terminal resistance of our device $R_{2T}$. For convenience and better visibility we use $R_{2T}$ (rather than the four-terminal resistance $R_{xx}$) in this color map; however, all the quantitative analysis later on will be performed using quantities not affected by the contact resistance, i.e. $\sigma_{xx}$, $\sigma_{xy}$ and $R_{xx}$. In this graph, we can conveniently identify different regions in the field-gate space in which the device becomes strongly resistive. These regions disperse as $t=0$, i.e. as vertical lines in the Wannier diagram, and appear to be strongly modulated by the $|t|=1$ gaps, which enclose regions of Landau filling $< 1$. The first insulating area at $n/n_0=0$ (reaching a maximum two terminal resistance $R_{2T}^{max}= 8.4$ M$\Omega$, and four terminal resistance $R_{xx}^{max}= 2.6$ M$\Omega$, at $T=1.5$ K) extends over the whole field range and corresponds to half filling of the main $N=0$ LL. A second insulating state ($R_{2T}^{max}= 0.5$ M$\Omega$, $R_{xx}^{max}= 0.3$ M$\Omega$) is located at $n/n_0=1$ and develops for $B>22$ T, i.e. in $B_{eff}=B-22$ T $>0$ (green box). A third (weaker) one ($R_{2T}^{max}= 0.2$ M$\Omega$, $R_{xx}^{max}=$ 70 k$\Omega$) develops at $n/n_0=-2$ for $B<22$ T, i.e. for $B_{eff}<0$ (orange box). By comparison with Fig.\ref{One} (b), it is evident that all the three insulating states also correspond to changes in the sign of $R_{xy}$, marking the boundary between local electron-doped and hole-doped regions. However, not every change in the carrier sign corresponds to insulating regions: notably, the satellite NPs do not develop into $B$-induced insulating state due to overlap with the robust Landau gaps from the main neutrality point. On the other hand, the $B_{eff}$-induced insulating states do not coexist with any state from main QH fan, which makes them observable in our experiment. The competition with Landau gaps developing in the same field-density region evidently constrains the possibility to experimentally access the insulating states. In addition, our experiment reveals a pronounced electron-hole asymmetry in the HB (see Figure \ref{One} (b)), clearly affecting the second and third insulating states, for which we did not find a particle-hole symmetric. This asymmetry was consistently seen in previous experiments on graphene-hBN superlattices \cite{Ponomarenko_13,Dean_13,Hunt_13} and reproduced  by calculations.\cite{Wallbank_13,Chen_14} The superlattice perturbation induced by an hexagonal substrate having inequivalent sublattice sites (i.e. hosting Boron and Nitrogen atoms) is considered responsible for this effect.\cite{DaSilva_15}

In the following, we use the notation ($\nu$,$\nu_L$)\cite{Yu_14} in order to label the incompressible states in the local fan diagrams and check the consistency of the two $B_{eff}$-induced insulating states within the Hofstadter picture. Here $\nu$ is the filling factor of the ‘parent’ QH state, while $\nu_L$ is the filling factor in the local Landau fan. The relations $t = \nu + \nu_L$ and $s = -\nu_L$ necessarily hold. Figure \ref{Two} (a) shows an enlarged view of the $n/n_0=1$, $B_{eff}>0$ region, centred on the (1,-1) insulating state. The index $\nu$ is given by the main QH state located at $n/n_0=1$ for $B=22$ T, i.e. $\nu=1$ (shaded area). This QH state determines the energy gap for the third generation Dirac particles at $\mathit{\Phi/\Phi_0}=1$, $n/n_0=1$. The local QH fan, resulting from occupation of single-degenerate local LLs, is shown in Fig.\ref{Two} (b) (along with the neighbouring one, which develops from the main $\nu=0$ state). The colour code corresponds to the expected value for the Hall conductivity, given by $\sigma_{xy}=t(e^2/h)$, (grey $=0$, blue $=-1$, red $=+1$, light blue $=-2$, light red $=+2$). Panel (c) shows line traces of $\sigma_{xy}$ and $\sigma_{xx}$ at $B=30$ T ($B_{eff}=8$ T), which matches the expectation of the local fan diagram in (b). Although this pattern was reproducible over several cool-downs, the accuracy in the quantization of $\sigma_{xx}$ and $\sigma_{xy}$ was found to vary as a result of thermal cycling (e.g., the state (0,1), which cannot be identified here, was close to quantization in a previous measurement session). As shown in panels (d-f), the same analysis is successfully applied to the (-2,2) insulating state in the $n/n_0=-2$, $B_{eff}<0$ region, which corresponds to a main filling factor $\nu=-2$, although the quantization at $B_{eff}=-7$ T is found to be generally less accurate than in the previous case.

\subsection{Energy gaps and microscopic origin}
\begin{figure}[!t]
\centerline{\includegraphics[width = 0.5\textwidth]{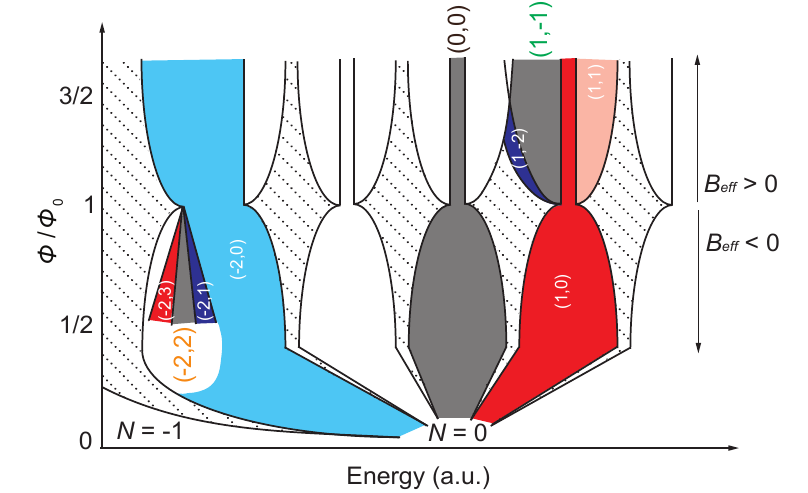}}
\caption{Schematic diagram for the evolution of the relevant energy levels and gaps with $\mathit{\Phi/\Phi_0}$ and $B_{eff}$. The superlattice-broadened $N=-1$ and (split) $N=0$ LLs are depicted as dashed areas, while the ($\nu$,$\nu_L$) states considered in our discussion are filled with a colour scale matching Fig.\ref{One} (b) and Fig.\ref{Two} (b,c) and (e,f).}
\label{Four}
\end{figure}

Having rationalised the presence of the (1,-1) and (-2,2) states in the graphene's HB, it is worthwhile to compare our observations with the existing experimental literature on the subject. In particular, Yu \textit{et al.} already identified both (1,-1) and (-2,2) states as incompressible in capacitance measurements,\cite{Yu_14} while Hunt \textit{et al.} showed vanishing two-terminal conductance both at (1,-1) and its particle-hole symmetric (-1,1).\cite{Hunt_13} Moreover, by carefully inspecting the colour plots in Ref.\onlinecite{Wang_15}, one can spot even a larger number of such highly resistive regions. However, quantitative information on the amplitude of the energy gaps and its connection to the physical origin of these states is still missing. Thereby, we measured $R_{xx}$($V_g$) in the vicinity of the three insulating states (0,0), (1,-1) and (-2,2), for increasing temperatures and different magnetic fields, with steps of 1 T. Typical $R_{xx}$($V_g$) traces for the three states, at representative values of $T$ and fixed $B$ ($B_{eff}$), are plotted in Fig.\ref{Three} (a-c). The insulating temperature dependence, i.e. the peak in $R_{xx}$ increasing with decreasing temperature and exceeding $h/e^2$, is evident in the three panels. We then analysed these data by fitting an Arrhenius type behaviour $R_{xx}$($T$)$=R_{0}\exp(\Delta/2k_BT)$ to the $T$-dependence at fixed values of $B$ ($\Delta$ is the energy gap and $k_B$ the Boltzmann constant), which is clearly emphasized in ln($R_{xx}$) vs 1/$T$ plots, as shown in Fig.\ref{Three} (d-f). Typically, this exponential dependence applies to relatively high temperature ranges, while $R_{xx}$ saturates at low $T$, where it results from hopping between localized states inside the energy gap. In Fig.\ref{Three} (g) we show the complete field dependence of the energy gaps of the three insulating states, obtained by fits of the kind shown by solid lines in panels (d-f); the error bars are given by the standard error in the fitting parameter $\Delta$. We discuss this panel by referring to the schematic diagram of Fig.\ref{Four}, which shows a simple model of Landau quantization for the third generation Dirac fermions and the resulting energy gaps.\cite{Chen_14,Yu_14} The main LLs are represented as dashed areas; in accordance to the experimental observations (see Fig.\ref{One} (b)), the main $N=0$ LL splits into four branches, while the $N=-1$ retains the four-fold degeneracy. The states relevant to our discussion are colour-filled, with a code intended to match the plots of Fig.\ref{Two}.
\\
The (0,0) insulating state extends over the entire field range considered in our experiment. Its energy gap can be estimated to be in the order of 150 K (14 meV) already at $B=1$ T (see open circles in Fig.\ref{Three} (g)). This energy scale is well beyond the single-particle spin splitting determined by the Zeeman energy $E_Z=g\mu_B B \approx 1.2$ [KT$^{-1}$]$\times B$ (where $g$ is the Land\'e factor, $\mu_B$ the Bohr magneton). It is instead comparable to the Coulomb energy $E_C=e^2 /4\pi\epsilon_0\epsilon_r l _B\approx 643/\epsilon_r$ [KT$^{-1/2}$]$\times\sqrt{B}$ (where $l_B$ is the magnetic length and $\epsilon_r$ the relative dielectric constant), which is plotted as a grey solid line in Fig.\ref{Three} (g), assuming $\epsilon_r=5$ for the non-encapsulated graphene-hBN sample used in this experiment.\cite{Dean_11} This observation is consistent with the formation of a spin-valley antiferromagnetic order at half filling of the $N=0$ LL,\cite{Kharitonov_12} in which electrons with opposite spin polarization occupy the two sublattice sites, minimizing the Coulomb repulsion.\cite{Young_14} The energy gap $\Delta(0,0)$ shows a markedly non-monotonic behavior as a function of magnetic field: it grows in the range 1 T $<B<16$ T (although significantly deviating from $E_C$ from 4 T on), it strongly decreases for 16 T $<B<22$ T, and finally increases again up to the highest fields applied. We attribute the deviation from $E_C$ to the fact that resistance data at relatively high temperature (up to $T=90$ K) were necessary to estimate $\Delta(0,0)$. At such temperatures the Zak mini-bands arising at rational values of flux quanta, recently identified in a new kind of quantum oscillatory phenomenon,\cite{Kumar_17,Kumar_18} compete with the thermally-excited conductivity across the gap, partially hindering the exponential dependence of $R_{xx}$. The dramatic suppression of the gap in the second region is due to the exponential broadening of the split LLs caused by the superlattice modulation, which reaches its maximum at $\mathit{\Phi/\Phi_0}=1$, where the band edges correspond to gapped Dirac cones.\cite{Wallbank_13} Local LL quantization in $B_{eff}>0$ yields to the opening of a $v_L=0$ state, which is reflected by the final increase in $\Delta$(0,0). These findings are fully consistent with previous magneto-capacitance spectroscopy measurements.\cite{Yu_14}\\
\begin{figure}[!t]
\centerline{\includegraphics[width = 0.43\textwidth]{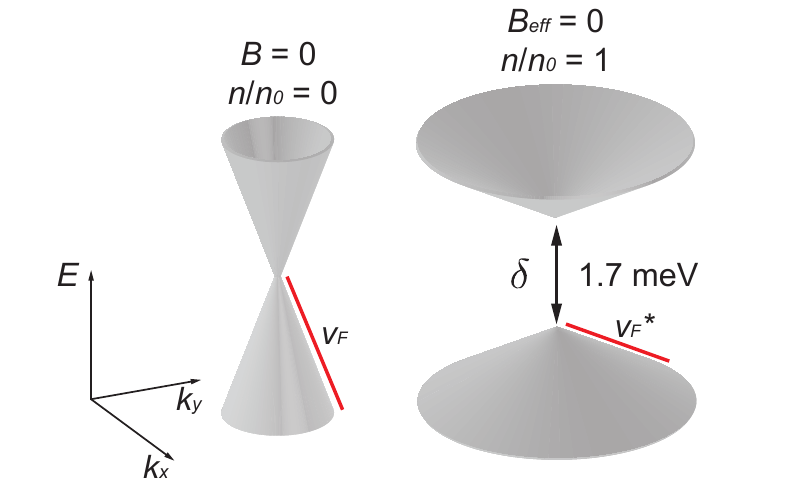}}
\caption{Comparison between the energy-momentum dispersion of Dirac fermions close to half filling in single-layer graphene (left), and the field-induced third generation ones at $\mathit{\Phi/\Phi_0}=1$, $n/n_0=1$ in a graphene superlattice (right). The cones' energy dispersion is given by $\pm v_F\hbar\sqrt{k_x^2+k_y^2}$ and $\pm[\delta/2+ v_F^*\hbar\sqrt{k_x^2+k_y^2}]$ respectively. The energy gap $\delta$ is determined by the $\nu=1$ broken-symmetry QH state at 22 T. The Fermi velocity $v_F^*=0.2\times v_F$ is quantitatively estimated by comparing the field-dependent energy gap of the insulating state (1,-1) with its analogous $\nu=-2$ in the main QH fan.}
\label{Five}
\end{figure}
The gap for the (1,-1) insulating state (open green triangles in Fig.\ref{Three} (g)) is sizeable for $B>27$ T ($B_{eff}>5$ T), increases monotonically with the field, and it is best fitted by a $\sqrt{B_{eff}}$-dependence. The parent quantum Hall state in this case is the $\nu=1$, (1,0) in the local notation. Its gap is known to be determined by the energy cost for the formation of skyrmionic spin textures,\cite{Young_12} which is proportional to $E_C$. The $\nu=1$ is found to extend continuously from $B=3$ T (see Fig.\ref{One} (b)), without experiencing any gap-closing, which was instead reported at $\mathit{\Phi/\Phi_0}=1$ in Ref.\onlinecite{Yu_14}. The fact that (1,0) remains gapped at $B_{eff}=0$ (with $\delta$(1,0)($B=22$ T) $\approx20$ K) is attributed to the smaller dielectric constant in our non-encapsulated sample, which enhances the interaction effects in comparison to the case of fully encapsulated structures ($\epsilon_r=8$), like the capacitance device use in Ref.\onlinecite{Yu_14}. On the other hand, the (1,-1) state results from full depletion of a single-degenerate $N=0$ - like local Landau level. Thereby, it is safe to assume that this insulating state has a single-particle origin, i.e. that it corresponds to the first cyclotron gap in the local single-degenerate Landau fan (see corresponding grey area in Fig.\ref{Four}), which amplitude is given by $v_F^*\sqrt{2\hbar e B_{eff}}$. Our best fit $\Delta(1,-1)=(15.7\pm0.8)$[KT$^{-1/2}$]$\times\sqrt{B_{eff}}$ can therefore provide a way to estimate the Fermi velocity of the corresponding third generation Dirac fermions $v_F^*$. However, it is well known that the experimental activation gap for the equivalent state in the conventional graphene spectrum ($\nu=-2$), although extremely large,\cite{Novoselov_07} highly underestimates the theoretical cyclotron gap. To circumvent this issue, we measured the activation gap of the (-2,0) state in our sample at $B=2$ T, where full quantization is already achieved and superlattice effects are minimized, and obtained $\Delta(-2,0)(B=2$ T) $=108$ K (theoretical value 590 K). We then estimated $v_F^*$ from the ratio $[\Delta(1,-1)/\sqrt{B_{eff}}]/[\Delta(-2,0)/\sqrt{B}]$. The direct comparison between two states with the same microscopic origin, i.e. first cyclotron gap in the Dirac spectrum, respectively in an effective and absolute field, is intended to take into account the sample-dependent localization mechanism that is responsible for the bulk gaps in the quantum Hall regime. Our analysis gives $v_F^*\approx 0.2\times v_F$, i.e. it indicates a significant renormalization of the band dispersion for the field-and-superlattice-induced Dirac fermions with respect to the ones of standard single-layer graphene ($v_F=10^6$ m/s). A schematic comparison is presented in Fig.\ref{Five}.\\
Finally, the field-dependence of the activation gap for the (-2,2) state is presented as open orange diamonds in Fig.\ref{Three} (g). This state appears to be less robust then the ones discussed above. This is reflected by its complete suppression at $\Phi/\Phi_0=1/2$, while the other insulating states remain sizeable at $\Phi/\Phi_0=3/2$ and beyond. The single (quasi) quantized steps of $\sigma_{xy}$ in the vicinity of (-2,2) (see Fig.\ref{Two} (f)) indicate that the degeneracy of the four-fold local $N=0$ LL for the replica Dirac fermions is fully lifted in negative $B_{eff}$(see Fig.\ref{Four}). The (-2,2) state corresponds to half-filling of this level, i.e. it is analogous to the (0,0) presented above. However, its gap is far from being comparable to $E_C$ and it is best fitted by $\Delta$(-2,2)$=(1.7\pm0.1)$[KT$^{-1}$]$\times B_{eff}$, which can be attributed to Zeeman splitting in an effective magnetic field, with an enhanced Land\'e factor $g^*\approx2.8$. This kind of $B$-dependence for the gap at half filling of the $N=0$ LL was typically reported for disordered graphene samples on SiO$_2$,\cite{Giesbers_09} while a much larger renormalized $g^*$ was estimated for graphene on hBN in Ref.\onlinecite{Young_12}. Despite the low-disorder environment guaranteed by the graphene/hBN stack, our observations indicate a moderate contribution of e-e interaction to the degeneracy lifting in this local Landau fan. Further experimental data would be however necessary to determine the exact microscopic ordering underlying the (-2,2) state; tilted-field experiments allowing for an independent tuning of the Zeeman field could be of particular relevance.

\section{Conclusions and Outlook}
In summary, we have presented a study of temperature-dependent magnetotransport on a graphene-hBN superlattice with 15 nm moir\'e periodicity. The electrical transport experiments allows the identification of three fully gapped regions in the HB. These states, despite sharing a common insulating nature, can be traced to different microscopic origin within the main QH spectrum of single-layer graphene and its replica in the vicinity of $\mathit{\Phi/\Phi_0}=1$ ($B=22$ T). Importantly, our analysis identifies the insulating state (1,-1) as corresponding to the first cyclotron gap in the local Landau fan of the replica Dirac fermions. The $B_{eff}$-dependence of its gap, in combination with knowledge of the $\nu=1$ gap at 22 T, enables to experimentally determine the energy-momentum dispersion of the corresponding superlattice-and-field induced third generation quasi-particles. An extension of this quantitative approach to samples with different moir\'e length should elucidate the role of the superlattice periodicity on the renormalization of the replica spectrum. Continuous tuning of the graphene-hBN misalignment via the method of Ref.\onlinecite{Palau_18} could be used for the creation of Dirac particles with on-demand Fermi velocity and gap size.

\section{Acknowledgments}
This work is part of the research programme no. 132 “High Field Magnet Laboratory: a global player in science in high magnetic fields”, financed by the Netherlands Organisation for Scientific Research (NWO). A.M. acknowledges the support of EPSRC Early Career Fellowship EP/N007131/1.

\end{document}